\documentclass[reprint,aps,prl,superscriptaddress,preprintnumbers,amssymb,amsmath,alignedleftspaceno]{revtex4-2} 
\pdfoutput=1
\usepackage[utf8]{inputenc}
\usepackage[T1]{fontenc}
\usepackage[USenglish]{babel}
\usepackage[group-digits=integer,quantity-product=\ ,uncertainty-descriptors={stat, syst}]{siunitx}
\usepackage{csquotes}
\usepackage[all]{foreign}
\defasforeign[ansaetze]{ansätze}
\renewcommand{\cf}{\xperiodafter{{\foreignabbrfont{cf}}}}

\usepackage{tikz}
\usepackage{braket}

\newcommand{\pd}[2]
{
    \frac{\partial #1}{\partial #2}
}
\usepackage[pdfusetitle,pdfauthor={Miguel Salg},bookmarksopen=true,bookmarksopenlevel=1]{hyperref}
\usepackage[english]{cleveref}
\usepackage{graphicx}
\usepackage{tikz}
\usepackage{longtable}
\usepackage[protrusion=true,expansion,tracking,kerning]{microtype}
\microtypecontext{spacing=nonfrench}
\newdimen\defaultaddspace
\begin{document}
\title{Precision calculation of the electromagnetic radii of the proton and \texorpdfstring{neutron\\}{neutron }from lattice QCD}
\author{Dalibor Djukanovic}
 \affiliation{Helmholtz Institute Mainz, Staudingerweg 18, 55128 Mainz, Germany}
 \affiliation{GSI Helmholtzzentrum für Schwerionenforschung, 64291 Darmstadt, Germany}
\author{Georg von Hippel}
 \affiliation{\texorpdfstring{PRISMA${}^+$}{PRISMA+} Cluster of Excellence and Institute for Nuclear Physics, Johannes Gutenberg University Mainz, Johann-Joachim-Becher-Weg 45, 55128 Mainz, Germany}
\author{Harvey B. Meyer}
 \affiliation{Helmholtz Institute Mainz, Staudingerweg 18, 55128 Mainz, Germany}
 \affiliation{\texorpdfstring{PRISMA${}^+$}{PRISMA+} Cluster of Excellence and Institute for Nuclear Physics, Johannes Gutenberg University Mainz, Johann-Joachim-Becher-Weg 45, 55128 Mainz, Germany}
\author{Konstantin Ottnad}
 \affiliation{\texorpdfstring{PRISMA${}^+$}{PRISMA+} Cluster of Excellence and Institute for Nuclear Physics, Johannes Gutenberg University Mainz, Johann-Joachim-Becher-Weg 45, 55128 Mainz, Germany}
\author{Miguel Salg}
 \email{msalg@uni-mainz.de}
 \affiliation{\texorpdfstring{PRISMA${}^+$}{PRISMA+} Cluster of Excellence and Institute for Nuclear Physics, Johannes Gutenberg University Mainz, Johann-Joachim-Becher-Weg 45, 55128 Mainz, Germany}
\author{Hartmut Wittig}
 \affiliation{Helmholtz Institute Mainz, Staudingerweg 18, 55128 Mainz, Germany}
 \affiliation{\texorpdfstring{PRISMA${}^+$}{PRISMA+} Cluster of Excellence and Institute for Nuclear Physics, Johannes Gutenberg University Mainz, Johann-Joachim-Becher-Weg 45, 55128 Mainz, Germany}
\begin{abstract}
    We present lattice-QCD results for the electromagnetic form factors of the proton and neutron including both quark-connected and -disconnected contributions.
The parametrization of the $Q^2$-dependence of the form factors is combined with the extrapolation to the physical point.
In this way, we determine the electric and magnetic radii and the magnetic moments of the proton and neutron.
For the proton, we obtain at the physical pion mass and in the continuum and infinite-volume limit $\sqrt{\langle r_E^2 \rangle^p} = \qty{0.820(14)}{fm}$, $\sqrt{\langle r_M^2 \rangle^p} = \qty{0.8111(89)}{fm}$, and $\mu_M^p = \num{2.739(66)}$, where the errors include all systematics.

\end{abstract}
\date{\today}
\preprint{MITP-23-045}

\maketitle
\newpage
\subpdfbookmark{Introduction}{introduction}
\paragraph{Introduction.}
The so-called \enquote{proton radius puzzle}, \ie the observation of a
large tension in the proton's electric (charge) radius extracted either
from atomic spectroscopy data of muonic
hydrogen \cite{Pohl2010,Antognini2013} or, alternatively, from
corresponding measurements on electronic hydrogen \cite{Mohr2012} as
well as $ep$-scattering data \cite{Bernauer2014,Mihovilovic2021}, has
gripped the scientific community for more than 10 years and triggered
a vigorous research effort designed to explain the discrepancy.

Recent results determined from $ep$-scattering data collected by the
PRad experiment \cite{Xiong2019} and from atomic hydrogen
spectroscopy \cite{Beyer2017,Bezginov2019,Grinin2020} (with the
exception of Ref.\@ \cite{Fleurbaey2018}) point towards a smaller
electric radius, as favored by muonic hydrogen and dispersive
analyses of $ep$-scattering data \cite{Mergell1996, Belushkin2007, Lorenz2015, Hoferichter2016, Alarcon2020, Lin2021a,
Lin2021, Lin2022}. To allow for a more reliable and precise
determination of the proton's electromagnetic form factors from which
the radii are extracted, efforts are underway to extend
$ep$-scattering experiments to unprecedentedly small momentum
transfers \cite{Grieser2018, Gasparian2022, Suda2022}, which are
complemented by plans to perform high-precision measurements of
$\mu p$ cross sections \cite{Cline2021,Quintans2022}.

While the situation regarding the electric radius is awaiting
its final resolution, one also finds discrepant results for the
proton's magnetic radius. Specifically, there is a tension of
$2.7\,\sigma$ between the value extracted from the A1 $ep$-scattering
data alone and the estimate from the corresponding analysis applied to
the remaining world data \cite{Lee2015}. Clearly, a firm theoretical
prediction for basic properties of the proton and the neutron, such as
their radii and magnetic moments, would be highly desirable in
order to assess our understanding of the particles that make up the
largest fraction of the visible mass in the universe.

In this letter we present our results for the radii and
magnetic moment of the proton computed in lattice QCD. Compared with
previous lattice studies \cite{Goeckeler2005, Yamazaki2009,
Syritsyn2010, Bratt2010, Alexandrou2013, Bhattacharya2014,
Shanahan2014a, Shanahan2014, Green2014, Capitani2015, Alexandrou2017,
Hasan2018, Ishikawa2018, Shintani2019, Alexandrou2019, Alexandrou2020,
Djukanovic2021, Ishikawa2021}, our calculation is the first to include
the contributions from quark-disconnected diagrams while controlling
all sources of systematic uncertainties arising from excited-state
contributions, finite-volume effects and the continuum extrapolation.
We determine the proton's magnetic radius
$\sqrt{\langle r_M^2 \rangle^p}$ with a total precision of \qty{1.1}{\percent},
which is competitive with recent analyses of $ep$-scattering
data \cite{Bernauer2014,Lee2015,Alarcon2020,Lin2021a}. Moreover, our lattice QCD
estimate for the proton's magnetic moment is in good agreement with
experiment. Our result for the electric radius, which has a
total precision of \qty{1.7}{\percent}, is consistent with the value
determined in muonic hydrogen within 1.5 standard deviations.

\currentpdfbookmark{Lattice setup}{setup}
\paragraph{Lattice setup.}
Our aim is to compute the electric and magnetic Sachs form factors $G_E(Q^2)$ and $G_M(Q^2)$ of the proton and neutron.
The electric form factor at zero momentum transfer yields the
nucleon's electric charge, \ie $G_E^p(0) = 1$ and $G_E^n(0) = 0$,
whereas the magnetic form factor at $Q^2=0$ is
identified with the magnetic moment, $G_M(0) = \mu_M$.
The radii can in turn be extracted from the slope of the form factors at zero momentum transfer,
\begin{equation}
    \langle r^2 \rangle = -\frac{6}{G(0)} \left.\pd{G(Q^2)}{Q^2}\right|_{Q^2 = 0} .
    \label{eq:radii}
\end{equation}
The only exception to this definition is the electric radius of
the neutron, where the normalization factor $1/G(0)$ is dropped.

For our lattice determination of these quantities, we use the
ensembles generated by the Coordinated Lattice Simulations (CLS) \cite{Bruno2015} effort with $2 + 1$ flavors of non-perturbatively $\mathcal{O}(a)$-improved Wilson fermions \cite{Sheikholeslami1985,Bulava2013} and a tree-level improved Lüscher-Weisz gauge action \cite{Luescher1985}, correcting for the treatment of the strange quark determinant using the procedure outlined in Ref.\@ \cite{Mohler2020}.
\Cref{tab:ensembles} shows the set of ensembles entering the analysis:
they cover four lattice spacings in the range from \qty{0.050}{fm} to \qty{0.086}{fm}, and several pion masses, including one slightly below the physical value (E250).
Further details on our setup of the simulations and the measurements of the two- and three-point functions of the nucleon can be found in the accompanying paper \cite{Djukanovic2023}.

\begin{table}[htb]
    \caption{Overview of the ensembles used in this study. Further details are contained in table~I of the accompanying paper \cite{Djukanovic2023}.}
    \label{tab:ensembles}
    \begin{ruledtabular}
        \begin{tabular}{lccccc}
            ID                   & $\beta$ & $t_0^\mathrm{sym}/a^2$ & $T/a$ & $L/a$ & $M_\pi$ [MeV] \\ \hline
            C101                 & 3.40    & 2.860(11)              & 96    & 48    & 227           \\
            N101\footnotemark[1] & 3.40    & 2.860(11)              & 128   & 48    & 283           \\
            H105\footnotemark[1] & 3.40    & 2.860(11)              & 96    & 32    & 283           \\[\defaultaddspace]
            D450                 & 3.46    & 3.659(16)              & 128   & 64    & 218           \\
            N451\footnotemark[1] & 3.46    & 3.659(16)              & 128   & 48    & 289           \\[\defaultaddspace]
            E250                 & 3.55    & 5.164(18)              & 192   & 96    & 130           \\
            D200                 & 3.55    & 5.164(18)              & 128   & 64    & 207           \\
            N200\footnotemark[1] & 3.55    & 5.164(18)              & 128   & 48    & 281           \\
            S201\footnotemark[1] & 3.55    & 5.164(18)              & 128   & 32    & 295           \\[\defaultaddspace]
            E300                 & 3.70    & 8.595(29)              & 192   & 96    & 176           \\
            J303                 & 3.70    & 8.595(29)              & 192   & 64    & 266
        \end{tabular}
    \end{ruledtabular}
    \footnotetext[1]{These ensembles are not used in the final fits but only to constrain discretization and finite-volume effects.}
\end{table}

To extract the effective form factors from two- and three-point correlation functions, we employ the ratio method \cite{Korzec2009} and the same estimators for the effective electric and magnetic Sachs form factors as in Ref.\@ \cite{Djukanovic2021}.
For further technical details, we again refer to the companion paper \cite{Djukanovic2023}.
The effective form factors are constructed for the isovector ($u - d$)
and the connected isoscalar ($u + d$) combinations, as well as for the
light and strange disconnected contributions.
In the isovector case, the disconnected contributions cancel.
The full isoscalar (octet) combination $u + d - 2s$, on the other hand, is obtained from the connected and disconnected pieces as
\begin{equation}
    G_{E, M}^{\mathrm{eff}, u+d-2s} = G_{E, M}^{\mathrm{eff, conn}, u+d} + 2G_{E, M}^{\mathrm{eff, disc}, l-s} .
    \label{eq:eff_ff_u+d-2s}
\end{equation}
Note that the disconnected part only requires the difference $l - s$
between the light and strange contributions, in which correlated noise
cancels and which can be computed efficiently by the one-end trick
\cite{McNeile2006,Boucaud2008,Giusti2019}.

We express all dimensionful quantities in units of the gradient flow time $t_0$ \cite{Luescher2010} using the determination of $t_0^\mathrm{sym}/a^2$ from Ref.\@ \cite{Bruno2017}.
Only in the final step, \ie after the extrapolation to the physical point, are the radii converted to physical units by means of the FLAG estimate of
\begin{equation}
    \sqrt{t_{0, \mathrm{phys}}} = \qty{0.14464(87)}{fm}
    \label{eq:sqrt_t0_phys}
\end{equation}
for $N_f = 2 + 1$ from Ref.\@ \cite{Aoki2021}.

\currentpdfbookmark{Excited-state anaylsis}{excited_states}
\paragraph{Excited-state analysis.}
\label{sec:excited_states}
Due to the strong exponential decay of the signal-to-noise ratio for baryonic correlation functions with increasing source-sink separation \cite{Hamber1983,Lepage1989}, an explicit treatment of the excited-state systematics is required in order to extract the ground-state form factors from the effective ones \cite{Ottnad2020}.
In this work, we employ the summation method \cite{Maiani1987,Doi2009,Capitani2012}.
It exploits the fact that the contributions from excited
states to effective form factors are parametrically more strongly
suppressed when the insertion of the electromagnetic current is summed
over timeslices between the source and sink.
In the asymptotic limit, the slope of the summed correlator ratio with
respect to the source-sink separation $t_\mathrm{sep}$ yields the ground-state form factor \cite{Capitani2015,Djukanovic2021}.

In our analysis, we monitor the stability of fit results for different starting values $t_\mathrm{sep}^\mathrm{min}$ of the source-sink separation.
Rather than selecting one particular value of $t_\mathrm{sep}^\mathrm{min}$ on each ensemble, we perform a weighted average over $t_\mathrm{sep}^\mathrm{min}$, where the weights are given by a smooth window function \cite{Djukanovic2022,Agadjanov2023} [\cf eq.~(18) in the accompanying paper \cite{Djukanovic2023}].

\begin{figure*}[htb]
    \includegraphics[width=\textwidth]{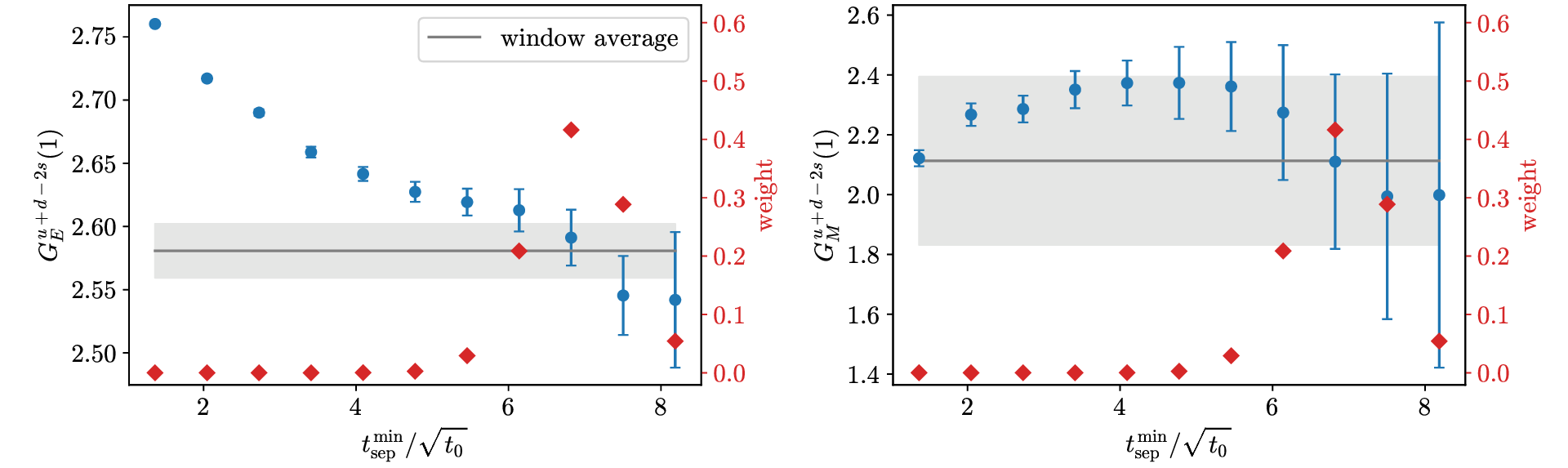} \caption{Isoscalar
    electromagnetic form factors at the lowest non-vanishing momentum
    ($Q^2 \approx \qty{0.067}{GeV^2}$) on ensemble E300 as a function
    of the minimal source-sink separation entering the summation
    fit. Each blue point corresponds to a single fit starting at the
    value given on the horizontal axis. The associated weights in the
    average are represented by the red diamonds, with the gray curves
    and bands depicting the averaged
    results.}  \label{fig:window_average}
\end{figure*}

This averaging strategy is illustrated in \cref{fig:window_average} for the isoscalar combination at the first non-vanishing momentum on ensemble E300.
One finds that the window averages agree within their errors with what one can identify as plateaux in the blue points.
This is observed for practically all other ensembles and momenta
employed in the analysis, and hence we conclude that the window method
reliably isolates the asymptotic value.
Moreover, it reduces the human bias arising from manually picking one particular value for $t_\mathrm{sep}^\mathrm{min}$ on each ensemble, because we use the same window parameters in units of $t_0$ on all ensembles.
Since our window average does not yield a significantly smaller error in comparison with the individual points entering the average (\cf \cref{fig:window_average}), we are confident that our error estimates are sufficiently conservative
to exclude any systematic bias in estimating the ground-state form factors.

Further crosschecks on our excited-state analysis, including a detailed comparison with an alternative approach based on two-state fits to the effective form factors, can be found in appendix~B of the accompanying paper \cite{Djukanovic2023}.
We did not find any evidence that our preferred strategy presented above introduces a systematic bias or underestimates the errors, but is on the contrary rather conservative in this regard.

\currentpdfbookmark{Direct BChPT fits}{bchpt_fits}
\paragraph{Direct B$\chi$PT fits.}
\label{sec:bchpt_fits}
To extract the radii from the form factors, we need to describe the $Q^2$-dependence of the latter.
As in Refs.\@ \cite{Capitani2015,Djukanovic2021}, we employ two different methods:
our preferred procedure is to combine the parametrization of the $Q^2$-dependence with the extrapolation to the physical point ($M_\pi = M_{\pi, \mathrm{phys}}$, $a = 0$, $L = \infty$) by fitting our form factor data directly to the expressions resulting from covariant baryon chiral perturbation theory (B$\chi$PT) \cite{Bauer2012}.
This is presented in the following.
Alternatively, we have implemented the more traditional strategy of first performing a generic parametrization of the $Q^2$-dependence on each ensemble, followed by extrapolating the resulting radii to the physical point.
A crosscheck of our main analysis with this two-step approach can be found in the accompanying paper \cite{Djukanovic2023}.

For our main analysis, we fit our form factor data to the full expressions of Ref.\@ \cite{Bauer2012} without explicit $\Delta$ degrees of freedom.
The fits are performed for the isovector and isoscalar channels separately, but for $G_E$ and $G_M$ simultaneously.
This allows us to properly treat the correlations between different $Q^2$ and also between $G_E$ and $G_M$.
Different gauge ensembles, on the other hand, are treated as statistically independent.
In the isovector channel, we include the contributions from the $\rho$ meson in the expressions for the form factors, while in the isoscalar channel, we include the leading-order terms from the $\omega$ and $\phi$ resonances.
The physical pion mass $M_{\pi, \mathrm{phys}}$ is fixed in units of $\sqrt{t_0}$ using its value in the isospin limit \cite{Aoki2014},
\begin{equation}
    M_{\pi, \mathrm{phys}} = M_{\pi, \mathrm{iso}} = \qty{134.8(3)}{MeV} ,
    \label{eq:m_pi_phys}
\end{equation}
\ie we employ $\sqrt{t_{0, \mathrm{phys}}} M_{\pi, \mathrm{phys}} = 0.09881(59)$.
Here, we neglect the uncertainty of $M_{\pi, \mathrm{iso}}$
in \unit{MeV} since it is completely subdominant compared to that of
$\sqrt{t_{0, \mathrm{phys}}}$ which enters in the conversion of units.

We perform several such fits with various cuts in the pion mass ($M_\pi \leq \qty{0.23}{GeV}$ and $M_\pi \leq \qty{0.27}{GeV}$) and the momentum transfer ($Q^2 \leq \qtyrange[range-phrase = {, \ldots, }, range-units = single]{0.3}{0.6}{GeV^2}$), as well as with different models for the lattice-spacing and/or finite-volume dependence, in order to estimate the corresponding systematic uncertainties.
The variations of the results due to the cuts are in most cases much smaller than their statistical errors and will be included in our quoted systematic errors.
We reconstruct the proton and neutron form factors as linear combinations of the B$\chi$PT formulae for the isovector and isoscalar channels, evaluating the low-energy constants as determined from the separate fits in these channels.
For further technical details of the fits, we refer to the accompanying paper \cite{Djukanovic2023}.

The major benefits of the direct approach compared to the two-step procedure have been previously observed in our publication on the isovector electromagnetic form factors \cite{Djukanovic2021}:
Firstly, including multiple ensembles as well as $G_E$ and $G_M$ in one fit significantly reduces the resulting errors on the radii.
Secondly, it greatly increases the number of degrees of freedom in the fit, which has a stabilizing effect with regard to lowering the momentum cut.

\currentpdfbookmark{Model average and final results}{model_average}
\paragraph{Model average and final results.}
\label{sec:model_average}
Since we do not have a strong \apriori preference for one specific setup of the direct fits, we determine our final results and total errors from averages over different fit models and kinematic cuts.
For this purpose, we use weights derived from the Akaike Information Criterion \cite{Akaike1973,Akaike1974,Neil2022,Burnham2004,Borsanyi2015}.
In order to estimate the statistical and systematic uncertainties of our model averages, we adopt a bootstrapped variant of the method from Ref.\@ \cite{Borsanyi2021}.
Our procedure is explained in more detail in the accompanying paper \cite{Djukanovic2023}.
As our final results, we obtain
{
    \allowdisplaybreaks
    \begin{align}
        \label{eq:rE2_p_final}
        \langle r_E^2 \rangle^p &= \qty{0.672(14)(18)}{fm^2} , \\
        \label{eq:rM2_p_final}
        \langle r_M^2 \rangle^p &= \qty{0.658(12)(8)}{fm^2} , \\
        \label{eq:muM_p_final}
        \mu_M^p                 &= \num{2.739(63)(18)} , \\[\defaultaddspace]
        \label{eq:rE2_n_final}
        \langle r_E^2 \rangle^n &= \qty{-0.115(13)(7)}{fm^2} , \\
        \label{eq:rM2_n_final}
        \langle r_M^2 \rangle^n &= \qty{0.667(11)(16)}{fm^2} , \\
        \label{eq:muM_n_final}
        \mu_M^n                 &= \num{-1.893(39)(58)} .
    \end{align}
}%
We note that the precision of the magnetic radius of the proton, $\sqrt{\langle r_M^2 \rangle^p} = \qty{0.8111(74)(50)}{fm}$, is commensurate with that of its electric counterpart, $\sqrt{\langle r_E^2 \rangle^p} = \qty{0.820(9)(11)}{fm}$.

\begin{figure*}[htb]
    \includegraphics[width=\textwidth]{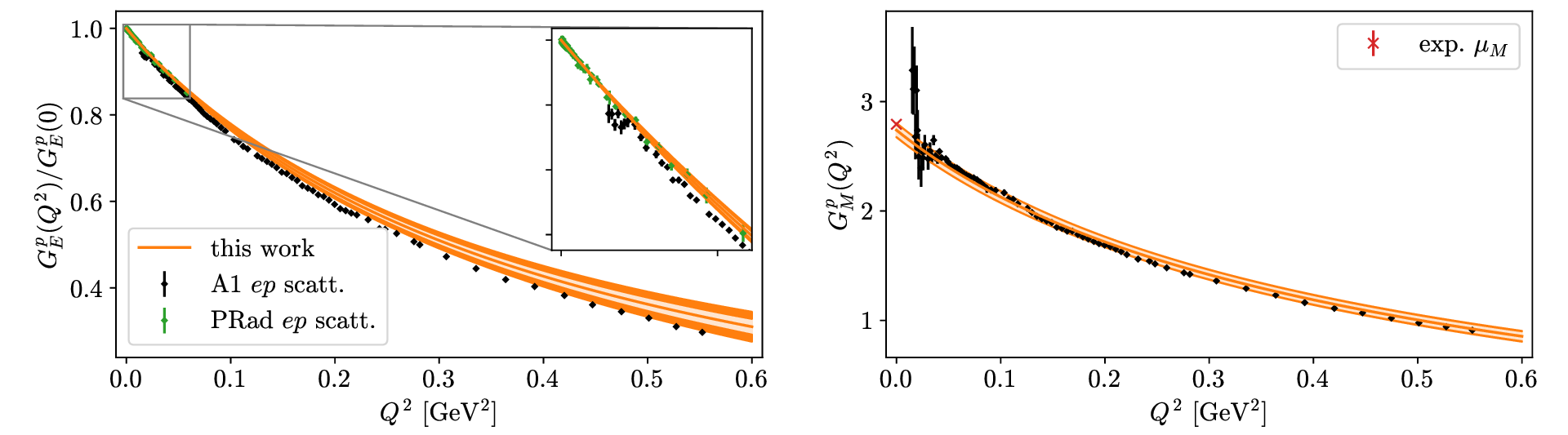}
    \caption{Electromagnetic form factors of the proton as a function
    of $Q^2$. The orange curves and bands correspond to our final
    results at the physical point with their full uncertainties
    obtained as model averages over the different direct fits. The
    light orange bands indicate the statistical errors. The black
    diamonds represent the experimental $ep$-scattering data by the
    A1 Collaboration \cite{Bernauer2014} obtained using Rosenbluth
    separation, and the green diamonds the corresponding data by PRad
    \cite{Xiong2019}. The experimental value of the magnetic
    moment \cite{Workman2022} is depicted by a red cross.}
    \label{fig:bchpt_fits_model_average}
\end{figure*}

To further compare our results to experiment we perform model averages of the form factors themselves.
The results are plotted in \cref{fig:bchpt_fits_model_average} for the proton.
One observes that the slope of the electric form factor as obtained from our calculation is closer to the PRad measurement \cite{Xiong2019} than to that of the A1 Collaboration \cite{Bernauer2014}.
The magnetic form factor, on the other hand, agrees well with the A1 data.
Moreover, our estimates reproduce within their errors the experimental results for the magnetic moments both of the proton and of the neutron \cite{Workman2022}.
The plots for the neutron corresponding to \cref{fig:bchpt_fits_model_average} in this letter are contained in fig.~7 of the accompanying paper \cite{Djukanovic2023}.

\begin{figure}[htb]
    \includegraphics[width=\linewidth]{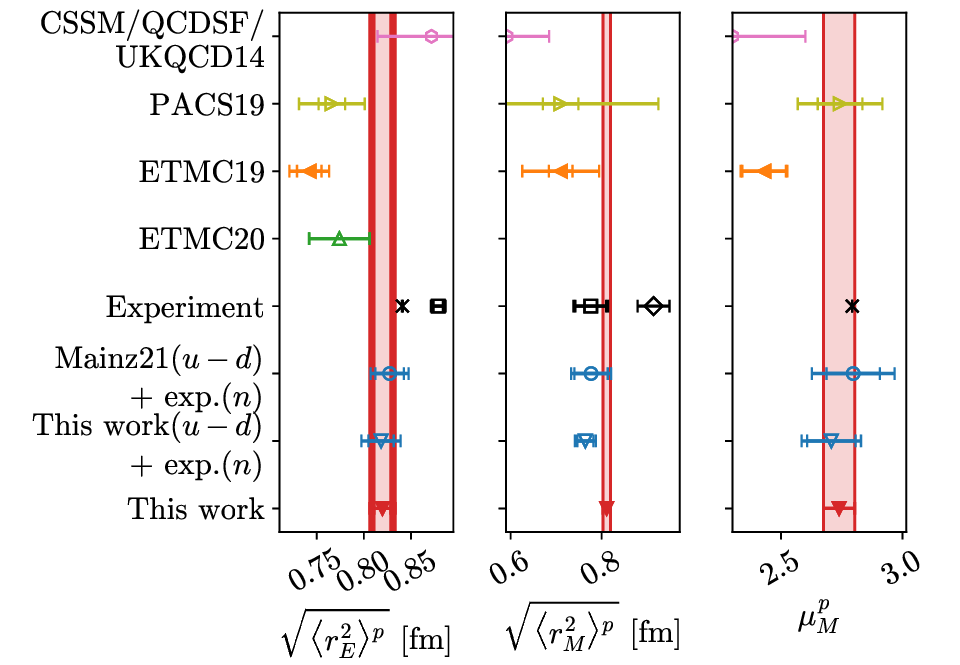}
    \caption{Comparison of our best estimates for the electromagnetic radii and the magnetic moment of the proton with other lattice calculations, \ie Mainz21 \cite{Djukanovic2021}, ETMC20 \cite{Alexandrou2020}, ETMC19 \cite{Alexandrou2019}, PACS19 \cite{Shintani2019}, and CSSM/QCDSF/UKQCD14 \cite{Shanahan2014a,Shanahan2014}. Only ETMC19 and this work include disconnected contributions. The Mainz21 values have been obtained by combining their isovector results with the PDG values for the neutron \cite{Workman2022}. We also show this estimate using our updated isovector results (\cf the accompanying paper \cite{Djukanovic2023}). The experimental value for $\mu_M^p$ is taken from PDG \cite{Workman2022}. The two data points for $\sqrt{\langle r_E^2 \rangle^p}$ depict the values from PDG \cite{Workman2022} (cross) and Mainz/A1 \cite{Bernauer2014} (square), respectively. The two data points for $\sqrt{\langle r_M^2 \rangle^p}$, on the other hand, show the reanalysis of Ref.\@ \cite{Lee2015} either using the world data excluding that of Ref.\@ \cite{Bernauer2014} (diamond) or using only that of Ref.\@ \cite{Bernauer2014} (square).}
    \label{fig:comparison}
\end{figure}

In \cref{fig:comparison}, our results for the electromagnetic radii and magnetic moment of the proton are compared to recent lattice determinations and to the experimental values.
We note that the only other lattice result including disconnected
contributions is ETMC19 \cite{Alexandrou2019}, which, however, has not
been extrapolated to the continuum and infinite-volume limits.
Our estimate for the electric radius is larger than the results
of Refs.\@ \cite{Alexandrou2019,Alexandrou2020,Shintani2019}, while
Ref.\@ \cite{Shanahan2014} quotes an even larger central value.

We stress that any difference between our estimate and previous
lattice calculations is not related to our preference for direct fits
to the form factors over the conventional approach via the
$z$-expansion, as the latter yields consistent values for the radii (\cf the accompanying paper \cite{Djukanovic2023}).
For the magnetic radius, our result agrees with that of Refs.\@ \cite{Alexandrou2019,Shintani2019} within $1.2$ combined standard deviations, while that of Ref.\@ \cite{Shanahan2014a} is much smaller.
Our statistical and systematic error estimates for the electric radius
and magnetic moment are similar or smaller compared to other lattice
studies, while being substantially smaller for the magnetic radius.
As a final remark we note that the lack of a data point at $Q^2 = 0$ complicates the extraction of the magnetic low-$Q^2$ observables in most recent lattice determinations, especially for $z$-expansion fits on individual ensembles.
By contrast, the direct approach -- in addition to combining
information from several ensembles and from $G_E$ and $G_M$ -- is more
constraining at low $Q^2$, allowing for considerably less variation in
the form factors in that regime.
We believe this to be responsible, to a large extent, for the small errors we achieve in the magnetic radii.

\currentpdfbookmark{Conclusions}{conclusions}
\paragraph{Conclusions.}
We have performed the first lattice QCD calculation of the
radii and magnetic moment of the proton to include the
contributions from quark-connected and -disconnected diagrams and
present a full error budget.
The overall precision of our calculation is sufficient to make a meaningful contribution to the debate surrounding the proton radii.
Our final estimates are listed in \cref{eq:rE2_p_final,eq:rM2_p_final,eq:muM_p_final,eq:rE2_n_final,eq:rM2_n_final,eq:muM_n_final}.

As an important benchmark, we reproduce the experimentally very precisely known magnetic moments of the proton and neutron \cite{Workman2022} within our quoted uncertainties.
A detailed discussion of our results for the neutron radii can be found in the accompanying paper \cite{Djukanovic2023}.
Our result for the electric (charge) radius of the proton is much closer to the value inferred from muonic hydrogen spectroscopy \cite{Antognini2013} and the recent $ep$-scattering experiment by PRad \cite{Xiong2019} than to the A1 $ep$-scattering result \cite{Bernauer2014}.
For the magnetic radius, on the other hand, our estimate is well
compatible with the analyses \cite{Bernauer2014,Lee2015} of the A1 data and exhibits a
$2.8\,\sigma$ tension with the other collected world
data \cite{Lee2015}.
The analyses of combined A1+PRad data \cite{Lin2021a} and A1 data alone \cite{Alarcon2020}, based respectively on dispersive and dispersively improved fit \ansaetze, arrive at values of $\sqrt{\langle r_M^2 \rangle^p}$ that are significantly larger than the A1-data analysis \cite{Lee2015} and in tension with our result.
This could partly be due to unaccounted-for isospin-breaking effects.
The shape of the magnetic form factor determined in our
lattice calculation, however, agrees very well with the measurement by A1 \cite{Bernauer2014} over the
whole range of $Q^2$ under study.

Returning to the (electric) proton radius puzzle, our lattice results lend further support to the emerging consensus that the issue has essentially been settled \cite{Hammer2020,Tiesinga2021,Antognini2022}.
Meanwhile, the situation regarding the magnetic radius remains to be clarified.

\begin{acknowledgments}
    \paragraph{Acknowledgments.}
    This research is partly supported by the Deutsche Forschungsgemeinschaft (DFG, German Research Foundation) through project HI 2048/1-2 (project No.\@ 399400745) and through the Cluster of Excellence \enquote{Precision Physics, Fundamental Interactions and Structure of Matter} (PRISMA${}^+$ EXC 2118/1) funded within the German Excellence Strategy (project ID 39083149).
    Calculations for this project were partly performed on the HPC clusters \enquote{Clover} and \enquote{HIMster2} at the Helmholtz Institute Mainz.
    Other parts were conducted using the supercomputer \enquote{Mogon 2} offered by Johannes Gutenberg University Mainz (\url{https://hpc.uni-mainz.de}), which is a member of the AHRP (Alliance for High Performance Computing in Rhineland Palatinate, \url{https://www.ahrp.info}) and the Gauss Alliance e.V.
    The authors also gratefully acknowledge the John von Neumann Institute for Computing (NIC) and the Gauss Centre for Supercomputing e.V. (\url{https://www.gauss-centre.eu}) for funding this project by providing computing time on the GCS Supercomputer JUWELS at Jülich Supercomputing Centre (JSC) through projects CHMZ21, CHMZ36, NUCSTRUCLFL, and GCSNUCL2PT.

    Our programs use the QDP++ library \cite{Edwards2005} and deflated SAP+GCR solver from the openQCD package \cite{Luescher2013}, while the contractions have been explicitly checked using the Quark Contraction Tool \cite{Djukanovic2020}.
    We thank Simon Kuberski for providing the improved reweighting factors \cite{Kuberski2023} for the gauge ensembles used in our calculation.
    Moreover, we are grateful to our colleagues in the CLS initiative for sharing the gauge field configurations on which this work is based.
\end{acknowledgments}

\bibliography{literature.bib}
\end{document}